\documentstyle[12pt]{article}
\begin{document}
\newcommand{\be}{\begin{equation}}
\newcommand{\ee}{\end{equation}}
\begin{center}

{\bf Soft and Hard Jets in QCD }\\

\vspace{2mm}

I.M. Dremin\footnote{Email: dremin@lpi.ru}\\
\vspace{2mm}

{\it Lebedev Physical Institute, Moscow 119991, Russia}\\

\end{center}

\begin{abstract}
Multiplicity of sets of soft jets with energies ranging in some interval
is determined. The possible role of collective effects is discussed.
\end{abstract}


The phenomenon of jet emission is well known in QCD and firmly established in
experiment. The two-jet events in $e^+e^-$-annihilation provide us with 
unique possibility to measure jets with fixed energy equal to that of 
colliding particles. In all other cases we have to deal with sets of jets
with different energies.
In three-jet events and in any high-$p_t$ process the produced jets are somehow
distributed in their energies. E.g., energies of gluon jets in three-jet
process change from some lower limit determined by the requirement to
separate this process to any value defined by experimental choice. Therefore,
one has to deal with sets of jets with energies ranging in some interval.

Multiplicity of jets  at a given energy has been calculated in QCD, and results
agree quite well with experiment (see the reviews in \cite{dre1, dgar, koch}).
This can be done for sets of jets as well \cite{dre2}. The proper weights for 
jets with different energies in the set are provided by the QCD equations.
The collective effects due to strings pulled apart during the separation of 
jets and color screening are somehow accounted in QCD. They lead, e.g., to
different suppression of multiplicities between $q\bar q$ and $qg$ pairs of 
jets. By measuring the multiplicity of sets of soft jets, this effect can
become more pronounced.

To simplify the presentation, I consider gluodynamics (see also \cite{dhwa}).
If the probability to create $n$ particles\footnote{In what 
follows, we adopt the local parton-hadron duality hypothesis with 
no difference between the notions of particles and partons up 
to some irrelevant factor.} in a jet is 
denoted as $P_n$, the generating function $G$ is defined as
\be
G(z,y) = \sum_{n=0}^{\infty }P_n(y)(1+z)^{n},                    \label{3}
\ee
where $z$ is an auxiliary variable,
$y=\ln (p\Theta /Q_0 )=\ln (2Q/Q_{0})$ is the evolution parameter, defining
the energy scale, $p$ is the initial momentum, $\Theta $ is the angle of 
the divergence of the jet (jet opening angle), assumed here to be 
fixed, $Q$ is the jet virtuality,  $Q_{0}=$ const.

The gluodynamics equation for the generating function is written as
\be
dG/dy= \int_{0}^{1}dxK(x)\gamma _{0}^{2}[G(y+\ln x)G(y+\ln (1-x)) - 
G(y)],      \label{geq}
\ee
where
\begin{equation}
\gamma _{0}^{2} =\frac {6\alpha _S}{\pi } ,                \label{52}
\end{equation}
$\alpha _{S}$ is the coupling strength and the kernel $K(x)$ is
\begin{equation}
K(x) = \frac {1}{x} - (1-x)[2-x(1-x)] .    \label{53}
\end{equation}

The equation for mean multiplicities follows from eq. (\ref{geq}):
\be
\langle n(y)\rangle ^{'} =\int_0^1dx\gamma _{0}^{2}K(x)(\langle n(y+\ln x)\rangle 
+\langle n(y+\ln (1-x))\rangle -\langle n(y)\rangle ).  \label{nav}
\ee
 As follows from eq. (\ref{nav}), the first 
two terms in the brackets correspond to mean multiplicities of two subjets,
and their sum is larger than the third term denoting the mean multiplicity
of the initial jet. Therefore, the integrand is positive.
This does not contradict to the statement that for a given event the total 
multiplicity is a sum of multiplicities in the two subjets because the 
averages in eq. (\ref{nav}) are done at different energies. 

The scaling property of the fixed coupling QCD \cite{dhwa} allows to look for 
the solution of the equation (\ref{nav}) with 
\be
\langle n\rangle \propto \exp (\gamma y) \;\;\;\; (\gamma = const)  \label{ny}
\ee

The anomalous dimension $\gamma $ is determined from (\ref{nav}) as
\be
\gamma =\gamma _0^2\int_0^1dxK(x)(x^{\gamma }+(1-x)^{\gamma }-1)
=\gamma _0^2M_1(1,\gamma ),                                   \label{gam}
\ee
where
\be
M_1(z,\gamma )=\int_0^zdxK(x)(x^{\gamma }+(1-x)^{\gamma }-1). \label{m_1}
\ee

For small enough $\gamma $ and $\gamma _0$ one gets
\be
\gamma \approx \gamma _0(1-0.458\gamma _0+ 0.213\gamma _0^2).   \label{gam0}
\ee

Now, according to the above discussion we define soft jets as those with 
sum of energies of belonging to them particles less than some $x_0E$. First, 
consider $x_0$=const and small. Then 
we should choose the upper limit of integration in eq. (\ref{nav}) equal to
$x_0$. 

One gets from (\ref{nav}) the mean multiplicity of a set of soft jets
$\langle n_s\rangle $:
\be
\frac{\langle n_s\rangle}{\langle n\rangle}=
\frac{M_1(x_0,\gamma)}{M_1(1,\gamma )}.            \label{nii}
\ee
For small $x_0$ it is
\be
\frac{\langle n_s\rangle}{\langle n\rangle} \approx  \frac{\gamma_0^2}{\gamma ^2}
x_0^{\gamma }N_1(x_0, \gamma ),        \label{nsn}
\ee
\be
N_1(x_0, \gamma )=1-\gamma ^2x_0^{1-\gamma }-\frac{2\gamma }{1+\gamma }x_0+
\frac{\gamma ^2(3+\gamma )}{4}x_0^{2-\gamma }+\frac{3\gamma }{2+\gamma }x_0^2 -
\frac{\gamma ^2 (2+\gamma )}{3}x_0^{3-\gamma }.     \label{n1g}
\ee
Thus we have found the energy dependence of mean multiplicity of particles in 
a set of subjets with low energies $E_s\leq x_0E$. As expected for constant
$x_0$, it is the same as the energy dependence of the total multiplicity with 
a different factor in front of it. Namely this dependence should be 
checked first in experimental data. Imposed on one another, these figures 
should coincide up to a normalization factor (\ref{nii}). This would confirm 
universality of gluons in jets.
 
Quite interesting is the non-trivial dependence of the normalization factor 
in eq. (\ref{nii}) on the parameter $x_0$, which does not coincide simply with
$x_0^{\gamma }$. It reflects the structure of QCD kernel $K(x)$. The main 
dependence on the cut-off parameter $x_0$ is given for $x_0\ll 1$ by the 
factor $x_0^{\gamma }$ with the same power as in dependence of
total multiplicity on energy. This corresponds 
to subjets with the largest energy of the set. However, with increase of $x_0$, 
this dependence is modified according to eqs (\ref{nii})-(\ref{n1g}). 
The negative corrections become more important in eq. (\ref{n1g}). They are
induced by subjets with energies lower than $x_0E$. The decrease of the
normalization factor corresponds to diminishing role of very low energy
jets at higher initial energies. This should be also checked in experiment.

If plotted as a function of the maximum energy in a set of jets $\epsilon_m$, 
the mean multiplicity is
\be
\langle n_s\rangle \propto \epsilon _m^{\gamma }[1-
\gamma ^2\left (\frac {\epsilon _m}{E}\right )^{1-\gamma }-
\frac{2\gamma }{1+\gamma }\left (\frac {\epsilon _m}{E}\right )+
\frac{\gamma ^2(3+\gamma )}{4}\left (\frac {\epsilon _m}{E}\right )^{2-\gamma }+
\frac{3\gamma }{2+\gamma }\left (\frac {\epsilon _m}{E}\right )^2 -
\frac{\gamma ^2 (2+\gamma )}{3}\left (\frac {\epsilon _m}{E}\right )^{3-\gamma }].
\label{eps}
\ee
It reminds eq. (\ref{ny}) with the correction factor in the brackets.

This is the consequence of the scaling property of the fixed coupling QCD which
results in the jets selfsimilarity. 

In principle, other definitions of soft jets are possible with $x_0=x_0(E)$. 
Then one should solve the equation 
\be
\frac{d\langle n_s\rangle }{dE} =E^{\gamma -1}
\gamma _{0}^{2}M_1(x_0(E), \gamma ),  \label{navs}
\ee
which follows from eq. (\ref{nav}). For example, one can choose the jets 
with energies less than some fixed constant independent of the initial 
energy. This would imply $\epsilon _m=$const or $x_0(E)\propto 1/E$, and the exact integration of 
eq. (\ref{navs}) is necessary. However, for qualitative estimates, eqs 
(\ref{nsn})-(\ref{eps}) can be used. They show that the ratio of average 
multiplicities (\ref{nsn})
tends to a constant at high energies corresponding to the multiplicity at 
the upper limit. At lower energies, it slightly increases with energy due to 
increasing role of jets with energies closest to their upper limit.

It is well known that for running coupling the power dependence $s^{\gamma /2}$
is replaced by $\exp(c\sqrt {\ln s})$. The qualitative statement about the 
similar energy behaviour of mean multiplicities in soft and inclusive processes 
should be valid also.

The above results can be confronted to experimental data if soft jets are 
separated in 3-jet events or in high-$p_t$ hadronic collisions. 
However, in our treatment we did not consider the 
common experimental cut-off which must be also taken into account.
This is the low-energy cut-off imposed on a soft jet for it to be
observable. It requires the soft jet not to be extremely soft. Otherwise the
third jet is not separated and the whole event is considered as a 2-jet one.
Thus the share of energy must be larger than some $x_1$, and the integration 
in eq. (\ref{m_1}) should be from $x_1$ to $x_0$. For $x_1\leq x_0\ll 1$ one
gets
\be
\frac{\langle n_s\rangle}{\langle n\rangle}=
\frac{\gamma_0^2}{\gamma ^2}[v(x_0)-v(x_1)],       \label{vx}
\ee
where the function $v(x)$ is easily guessed from eqs (\ref{nsn}), (\ref{n1g}).
At $x_1\ll x_0\ll 1$ eq. (\ref{nsn}) is restored. 

The values for hard jets are obtained by subtracting these results from 
values for the total process. 

I concentrated here on mean multiplicity but similar calculations have 
been done \cite{dre2} for higher moments of multiplicity distributions,
and they can be compared with experiment.

The main problem in comparison is related to the jet definition, i.e., to
treatment of particles belonging to the regions between the jets.  They are
more influenced by the collective effects which appear even in $e^+e^-$
annihilation due to strings pulled between jets and their mutual color 
screening. The experimental verification of the normalization factor in
(\ref{nii}) (for small $x_0$ it is the behavior of $N_1$ (\ref{nsn}))
becomes important because it would show that these collective effects
are properly accounted in QCD in a wide energy interval.

Another interesting experimental aspect of collective effects is the energy
distribution among jets in three-jet events. If boldly treated,
the QCD diagram of the process implies that one of the quark jets remains
untouched and must have the same energy as in two-jet events. Another
(anti)quark jet emits a gluon, and they share this energy. Therefore, the 
energy distribution should have the two-bump structure with one bump at the 
primary energy and another one at smaller values (near half of it if the 
energy is shared equally between quarks and gluons). This statement is correct 
if collective effects due to string tension play minor role. In principle, 
string tension can lead to some collective energy flow from one jet to 
another one and change the shape of the energy distribution among three 
produced jets. Thus one can determine the role of collective effects 
by measuring this distribution.

In conclusion, the experimentally measured values of mean multiplicities of 
particles belonging to a set of soft jets
can be compared with the obtained above theoretical predictions at different 
values of this share of energy. For a constant share, this dependence is
the same as for the average total multiplicity but with non-trivial 
$x_0$-dependence of the factor in front of it. Some predictions are obtained for 
energy dependent cut-offs. The collective effects due to string tension are
discussed. These results can be confronted to experiment. The results concerning
the behavior of higher moments analogous to those for fixed energy jets
\cite{dne} will be published elsewhere. \\

{\bf Acknowledgments}\\

This work has been supported in part by the RFBR grants 03-02-16134, 
04-02-16445, NSH-1936.2003.2.\\


\begin{thebibliography}{99}
\bibitem{dre1}
I.M. Dremin, UFN 164 (1994) 785; Physics-Uspekhi 37 (1994) 715.
\bibitem{dgar}
I.M. Dremin and J.W. Gary, Phys. Rep. 349 (2001) 301.
\bibitem{koch}
V.A.Khoze and W. Ochs, Int. J. Mod. Phys. A 12 (1997) 2949. 
\bibitem{dre2}
I.M. Dremin, JETP Lett. 81 (2005) 391.   
\bibitem{dhwa}
I.M. Dremin and R.C. Hwa, Phys. Lett. B 324 (1994) 477;
Phys. Rev. D 49 (1994) 5805.
\bibitem{dne}
I.M. Dremin, V.A. Nechitailo, Mod. Phys. Lett. A  9 (1994) 1471;
JETP Lett. 58 (1993) 881.
\end{thebibliography}
\end{document}